\documentclass[12pt]{article}
\usepackage{graphicx,amssymb}
\newcommand{\be}{\begin{equation}}
\newcommand{\ee}{\end{equation}}
\newcommand{\ba}{\begin{eqnarray}}
\newcommand{\ea}{\end{eqnarray}}

\newcommand{\n}{\label}

\begin{document}

\title{The Phantom of the OPRA}

\author{ \large A. Coley, S. Hervik and  J. Latta\\
{\it  \small Dalhousie University, Dept. of Mathematics and Statistics,}\\
{\it  \small Halifax, NS, Canada B3H 3J5,}\\
{\tt\small aac,~ herviks,~ lattaj@mathstat.dal.ca}\\}
\date{\small\today}
\maketitle
\begin{abstract}

The Observed Perlmutter-Riess Acceleration (OPRA)
implies that the expansion of the Universe is currently
increasing and is motivation for the so-called phantom energy models. 
We consider the dynamics of phantom scalar field models. 
An important physical time constraint, which
can be used to rule out many cosmological models, 
is obtained from the
condition that all forms of energy density, including the field
causing OPRA
(e.g., the phantom field), must be non-negligible for an extended
period, which is conservatively estimated to be of the order of a few $Gyr$.
We find that this physical time constraint cannot be satisfied in 
conventional phantom cosmological models.

\end{abstract}

\section{ACT I: OPRA}

Evidence from supernovae observations \cite{OPRA} strongly 
suggest the possibility of late-time accelerated expansion of the
Universe; the Observed  Perlmutter-Riess Acceleration (OPRA).
The energy density is usually described by an effective ``equation-of-state''
parameter $\gamma_{eff} - 1\equiv p/\rho$, the ratio of the energy pressure
$p$ to its energy density $\rho$.  A value $\gamma_{eff}<2/3$ is required
for cosmic acceleration.
The most recent observations, including supernovae, cosmic
microwave background experiments, 
quasar-lensing statistics and galaxy clustering  observations  (see
\cite{caldwell} and references therein), taken together
suggest a Universe with $\gamma_{eff}$ satisfying $\gamma_{eff} \le 0$; this
is consistent with $\gamma_{eff} = 0$, but it has
been argued that  with certain prior assumptions the best fit is
actually for $\gamma_{eff}<0$ \cite{caldwell,CKW}. It is therefore important
to look for theoretical possibilities to describe dark energy with
$\gamma_{eff} < 0$.

A wide class of scalar field cosmologies have
been utilized to model dark energy. Quintessence models
\cite{quint} lead to $0<\gamma_{eff}<2/3$. The simplest
explanation for dark energy is a cosmological constant, for which
$\gamma_{eff} = 0$. Matter with $\gamma_{eff}<0$ has been dubbed ``phantom energy''
\cite{caldwell}, and has received increased attention
recently \cite{more}. Specific models with non-minimally coupled
scalar fields may lead to phantom energy; perhaps the simplest
alternative is provided by a phantom scalar field with negative
kinetic energy. Theorists have also discussed stringy phantom
energy \cite{string} and brane-world phantom energy
\cite{braneworlds}.

Models with a constant $\gamma_{eff} < 0$ lead to a future singularity  commonly
called the {\it Big Rip}.  This singularity is typically
characterized by a divergent pressure and acceleration in a finite
proper time (although
there are rip-free phantom cosmologies and there are also sudden future singularities that are
less severe \cite{barrow}).
The energy density of phantom field increases and
eventually violates the dominant energy condition. 
The physical consequences in
this scenario have been discussed in \cite{CKW}. 
Expanding universes that come to a violent end after a finite
proper time at a future Big Rip singularity have a number of
additional undesirable properties \cite{Lima}. 

In this paper we take a conservative view and we will only
analyze the cosmological dynamics of the phantom field and relate
it to observations. The dynamics of phantom cosmology has been
discussed previously \cite{more,HaoLi}.

\section{ACT II: The Phantom}

 The Einstein equations in a
Robertson-Walker spacetime containing phantom matter and a separately conserved perfect
fluid  are
 \begin{equation}
 2\dot H=-(\rho_b+p_b+\rho_{ph}+p_{ph})+\frac{2k}{3a^{2}}
 \label{doth}
 \end{equation}
\begin{equation}
3H^2=(\rho_b+\rho_{ph}) - \frac{3k}{a^{2}}, \label{hubble}
\end{equation}
\begin{equation}
\dot{\rho_b}+3H(\rho_b +p_b) =0,  \label{con}
\end{equation}
where $a(t)$ is the expansion scale factor,
 $k$ is the curvature parameter, and  $H=\dot{a}/a$ is the Hubble expansion rate.
 In the above we have assumed that the
perfect fluid satisfies the barotropic equation of state
\begin{equation}
p_b = (\gamma -1) \rho_b, \label{6}
\end{equation}
where $\gamma$ is a constant which satisfies $0 < \gamma < 2$.

In the best motivated models the phantom cosmologies are generated by a (separately conserved) scalar field with
negative kinetic term. The energy density and pressure of
the field are given by
\begin{eqnarray}
\rho_{ph}=-\frac{1}{2}\dot\phi^2+V(\phi),\\
p_{ph}=-\frac{1}{2}\dot \phi^2-V(\phi).\end{eqnarray} The
corresponding equation of state parameter is now given by $(\gamma_{ph}-1)
\equiv {p_{\phi}}/{\rho_{\phi}}<0$, for $\rho_{\phi} >0$. The
effective Klein-Gordon equation reads \be \n{kg}
\ddot\phi+3H\dot\phi-\frac{dV}{d\phi}=0. \ee
We shall consider a  self-interacting scalar field with an
exponential potential 
\begin{equation}
V =V_0 e^{\kappa \phi}, \label{1}
\end{equation}
where $V_0$ and $\kappa$ are positive constants.
We note that a qualitative analysis of the standard spatially homogeneous and isotropic 
scalar field models has shown that there exist scaling
solutions which can act as unique late-time attractors in these models \cite{tracking}. 

\subsection{Dynamics}

Defining
\begin{equation}
\Phi \equiv \frac{\dot{\phi}}{\sqrt{6}H} \quad , \quad \Psi \equiv
\frac{V}{3H^2}\quad , \quad \Omega_b \equiv \frac{\rho_b}{3H^2},  \label{8}
\end{equation}
and the new logarithmic time variable $\tau$ by
\begin{equation}
\frac{d \tau}{dt} \equiv H, \label{9}
\end{equation}
the governing equations can be written as the three-dimensional
autonomous system:
\begin{eqnarray}
\Phi' &=&  \sqrt{\frac{3}{2}}\kappa \Psi + \frac{3}{2}\Phi \left[ \left(
\gamma - \frac{2}{3} \right) \Omega_b - \frac{2}{3} (2 + 2\Phi^2 + \Psi)
\right], \label{10}\\
\Psi' &=& 3 \Psi \left[ \sqrt{\frac{2}{3}} \kappa \Phi + \left(
\gamma -\frac{2}{3}  \right) \Omega_b - \frac{2}{3} (2\Phi^2 + \Psi - 1)
\right], \label{11}\\
\Omega_b' &=& 3 \Omega_b \left[\left(\gamma - \frac{2}{3}\right)(\Omega_b
-1) - \frac{2}{3} (2\Phi^2 + \Psi)  \right], \label{12}
\end{eqnarray}
where a prime denotes differentiation with respect to $\tau$, and the Friedmann 
equation becomes
\begin{equation}
1 - \Omega_b -  (\Psi - \Phi^2) = -k/a^2H^{2} \label{13}
\end{equation}

The physical region of the state-space is constrained by the
requirement that the energy densities be non-negative; i.e., $\Omega_b
\geq 0$ and $\Omega_{ph} \equiv (\Psi - \Phi^2) \ge 0$. 
Furthermore, from equation (\ref{13}) we find that the variables 
$\Omega_b$ and $\Omega_{ph}$
are bounded for $k=0$ and $k=-1$, but not for $k=+1$. The individual variables need
not be bounded. Note that $\Omega_{ph} = 0$ is {\em not} an invariant set.

Geometrically the zero-curvature models ($k=0$) are represented by 
$\Omega_b - \Phi^2 +\Psi = 1$,
in the $(\Phi,\Psi,\Omega_{b})$ state-space. Defining $\mathcal{K} = 
1 - \Omega_b - (\Psi - \Phi^2)$, we see that
\begin{equation}
\mathcal{K}' = \mathcal{K} \left[3\left(\gamma - \frac{2}{3}\right)\Omega_b
 -2(2\Phi^2 + \Psi)  \right],
\end{equation}
so that $\mathcal{K} = 0$ is an invariant set.

\subsubsection{Equilibrium points at finite values}

The equilibrium points $(\Phi_0,\Psi_0,\Omega_{b0})$ in the physical
phase space at finite values are:

\noindent
{\bf {\cal{M}}} $(0,0,0)$: eigenvalues $\left[-2, 2, (2-3\gamma)\right]$ ({\em saddle}). 
The exact solution corresponding to this equilibrium point is the Milne model.

\noindent
{\bf {\cal {F}}} $(0,0,1)$: eigenvalues $ \left[-\frac{3}{2}(2-\gamma)), 3\gamma, (3\gamma -2)\right]$ 
({\em saddle}). 
The exact solution corresponding to this equilibrium point is the 
flat Friedmann model.

\noindent
{\bf {\cal {R}}} $(\frac{\kappa}{\sqrt{6}}, 1 + \frac{\kappa^2}{6} ,0)$: 
eigenvalues 
$\left[ -(2 +\kappa^2), -3(1 + \frac{\kappa^2}{6}),
-(3\gamma +\kappa^2)\right]$ 
({\em sink}). 
The exact solution corresponds
to a {future late-time singularity} called the Big Rip singularity.
In this power-law
solution \cite{more}
\begin{equation}
a=a_0(-t)^{-2/\kappa^2}\label{a_phantom}, 
\end{equation}
\begin{equation}
V=\frac{2(6+\kappa^2)}{\kappa^4}\,e^{\kappa\phi},
\end{equation}
\begin{equation}
 \n{fi} \phi=\frac{-2}{\kappa}\ln{|t|}. 
\end{equation}
This solution represents a universe in which the scale factor
grows monotonically till a future Big Rip is reached at $t=0$. In
our models, $\rho>0$ but the weak energy condition is violated
because $\rho+p \sim -\dot\phi^2<0$. From (\ref{a_phantom}), 
the effective barotropic index for the
power-law phantom field solution is
$\gamma_{ph}=-{\dot\phi^2}/{\rho_{ph}}=-{\kappa^2}/{3}$
which becomes negative while both the energy density $\rho_{ph}$ and the
pressure $p_{ph}$ of the field diverge (at $t=0$).

We note that there are {\em no equilibrium points
corresponding to matter scaling solutions}, in which both
ordinary/dark matter and phantom matter are non-negligible $({\Phi_0}^2+\Psi_0) \ne 0,
\Omega_{b0} \ne 0$
\cite{tracking}.
This was noted in
\cite{HaoLi}. In addition, it was shown
\cite{Guoetal} that there are no matter scaling solutions
in two scalar field models (with exponential potentials).

\subsubsection{Equilibrium points at infinity}

We need to consider what happens if the variables diverge. This can occur if
$H=0$, thus leading to  bouncing models \cite{bouncing}. 
Let us therefore assume that $k\leq 0$ and $\Phi,\Psi \rightarrow \infty$. 
From the equations of motion we note that this can only happen to the past. 
In fact, the variables (including $\Omega_b$) will in general  
diverge to the past; asymptotically, we find
\begin{equation}
\Phi^2,~\Psi, ~\Omega_b \propto \frac{1}{\tau-\tau_0}, \quad \tau>\tau_0,
\end{equation}
and hence, the variables blow up in finite $\tau$. 
This blow-up is only an artifact of the variables chosen 
and results because the universe experiences a bounce. 
In principle, alternative dimensionless variables can be chosen 
to render the state space compact, even when the universe bounces. 
More details are presented in \cite{latta}.

A more serious problem with these models is the behaviour of the phantom energy density. 
We find that $\Omega_{ph}=\Psi - \Phi^2$ goes negative into
the past, and that the models, without modifications, are not physical \cite{latta}.

\subsection{Alternative Phantoms}

There is an alternative approach to obtaining Big Rip singularities
in which the phantom energy is modelled as a (separately conserved)
perfect fluid with effective equation of state
$p_{ph} = (\gamma_{ph} -1)\rho_{ph}$, where $\gamma_{ph}<0$ is a
constant. Defining normalized variables, all
equilibrium points are found to have either $\Omega_b$ or $\Omega_{ph}$
zero, and hence again there are no matter scaling solutions. Indeed, defining
$\chi = (\Omega_{ph} - \Omega_b)/(\Omega_{ph} + \Omega_b)$, we find that
$\chi$ is monotonically increasing from $\chi=-1$ to $\chi=+1$. These models cannot
be physical.

It has been noted that matter scaling solutions are not possible in phantom scalar field models 
with other potential \cite{HaoLi}. However, scaling solutions are possible if there 
are interactions allowed between the phantom field and the background
matter fields \cite{inter}; the physical consequences of these interacting models 
remain to be fully investigated.


\section{ACT III: the Phantom Unmasked?}
The physical models need to satisfy a number of observational constraints.
In general, the
Universe is approximately flat (but has non-zero curvature),
the Hubble time $H{_0}^{-1}$ corresponds to  $h = 0.71 \pm 0.06$
\cite{Spergel}, and it is strongly indicated 
that presently 5 percent of matter is due to baryons/ordinary matter, 
25 percent dark matter, 
and 70 percent dark energy. 
The coincidence problem (or fine tuning problem) asks 
why are the matter density, scalar density, and phantom density comparable at 
the present epoch.

In particular, the matter and scalar fields must be non-negligible
for an extended period in order to be consistent with galaxy
formation, the existence of dark matter
and OPRA. This is difficult to achieve in regular cosmology, although it can be 
accomplished in matter plus scalar field models due to the existence of
a matter scaling solution \cite{tracking},
which corresponds to an equilibrium point 
in a dynamical systems  analysis 
(but not necessarily an attractor; usually in  the presence
of curvature this equilibrium point is a saddle, and if trajectories move close
to this saddle, the trajectories will stay close for an arbitrarily long
period of time before  subsequently evolving away).

The transition from deceleration to acceleration occurs at redshift $z_T > 0.4$ 
\cite{padman2,Huterer}, and can be higher depending on the parameters of
the model; this redshift corresponds to
$t_T = 0.6 t_0$, where $t_0$ is the present time. Consequently, the time period during which the phantom field
must be non-negligible, ${\bf t}_{ph}$, satisfies
${\bf t}_{ph} > t_0 - t_T \sim \frac{1}{4} (H{_0}^{-1}) $.

In the theory of structure formation the 
present day galaxies and galaxy clusters formed due to
gravitational instability of initially small primordial density
fluctuations \cite{padman2}. Thus the time ${\bf t}_b$ for which the matter
is non-negligible must be at least as long as the 
age of globular clusters \cite{Spergel}: consequently, ${\bf t}_b> 12.0 ~ Gyr$.
We also note that ${\bf t}_{dm} > t_0 - t_{rec}$.

We shall not impose all of these constraints here, but simply consider the
condition that all forms of energy density must be non-negligible for an extended
period, which we very conservatively estimate to be ${\bf t} = \frac{\ell}{10} (H{_0}^{-1}) $,
where ${\ell}$ is of order unity, or a few $Gyr$.

\subsection{Discussion}

In the models we start with the phantom field energy density
 being dominated by the density of the ordinary matter. That is, we begin the
 evolution close to the equilibrium point {\bf {\cal {F}}}.
Eventually the energy density of the phantom field dynamics "switches on" (the 
precise details depending on the shape of the phantom field
potential $V(\phi)$ \cite{more}). The model then evolves towards the equilibrium point
{\bf {\cal {R}}}. 
\begin{figure}
\centering
\includegraphics{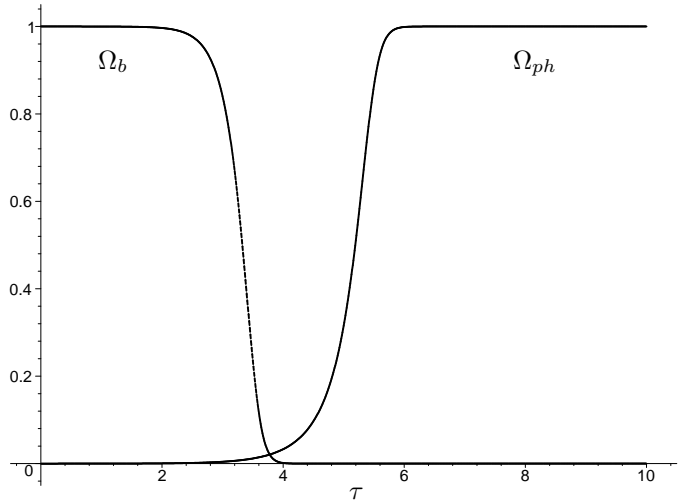}
\caption{The evolution of the phantom energy density, $\Omega_{ph}$, and the energy density of matter, $\Omega_b$, in terms of the logarithmic time $\tau$.} \label{Fig1}
\end{figure}

Numerical experiments were performed for different initial conditions. The results
presented in Figure 1  are typical. In Figure 1,  $\Omega_b$ and $\Omega_{ph}$ are 
only non-negligible for a very brief period. 
$\Omega_b$ changes
from unity to zero, as it moves away from the equilibrium point {\bf {\cal {F}}}
towards the sink {\bf {\cal {R}}}, at an exponential rate (similarly for
$\Omega_{ph}$). For all numerical values (and for all values of $k$)
we find that ${\bf t}$ is negligible, and the physical time constraint cannot 
be satisfied. We note from Figure 1 that during the epoch in which $\Omega_b$
and $\Omega_{ph}$ are non-negligible, the curvature is also non-negligible,
which is not consistent with current observations. By choosing particular initial
conditions (fine-tuning), we can arrange for the curvature to be small. But no
amount of fine-tuning can allow  for a significant ${\bf t}$. Hence these
models are not physical.  We emphasise that this {\em physical time constraint is
a very severe constraint and can be used to rule out many cosmological models}.

The behaviour to the past (which does depend on the sign of $k$)
was briefly discussed above. The numerics confirm that in all cases $\Omega_b$
crosses unity at a finite time in the past, whence $\Omega_{ph}$ becomes negative
and eventually the variables diverge. Presumably additional terms must be added to the
models to avoid this type of unphysical behaviour.

\section*{Acknowledgments}

This work was supported by NSERC (AC) and the Killam trust (SH).

\end{document}